\documentclass[11pt,reqno]{amsart}
\usepackage{a4wide}

\usepackage{amssymb, amsbsy, amsfonts}
\usepackage{amsmath, amsthm}

\newtheorem{Theorem}{Theorem}[section]
\newtheorem{Corollary}[Theorem]{Corollary}
\newtheorem{Lemma}[Theorem]{Lemma}
\theoremstyle{definition}
\newtheorem{Definition}[Theorem]{Definition}
\newtheorem{Example}[Theorem]{Example}
\theoremstyle{remark}
\newtheorem{Remark}[Theorem]{Remark}
\numberwithin{equation}{section}

\newcommand{\EE}{\mathbb{E}}
\newcommand{\ZZ}{\mathbb{Z}}
\newcommand{\QQ}{\mathbb{Q}}
\newcommand{\KK}{\mathbb{K}}
\newcommand{\FF}{\mathbb{F}}
\newcommand{\NN}{\mathbb{N}}
\newcommand{\RR}{\mathbb{A}}
\newcommand{\ssum}[2]{\text{\small$\displaystyle\sum_{#1}^{#2}$}}
\newcommand{\sprod}[2]{\text{\small$\displaystyle\prod_{#1}^{#2}$}}
\newcommand{\pisiSE}{$\Pi\Sigma^*$}
\newcommand{\sigmaSE}{$\Sigma^*$}
\newcommand{\piE}{$\Pi$}
\newcommand{\SigmaP}{{\sf Sigma}}
\newcommand{\notion}[1]{{\it #1}}
\newcommand{\dfield}[2]{({#1},{#2})}
\newcommand{\set}[2]{{\{#1\,|\,#2\}}}
\newcommand{\fct}[3]{{#1:#2 \to #3}}
\newcommand{\ev}{{\rm ev}}
\newcommand{\seqR}{{S}(\KK)}
\newcommand{\seqP}[1]{{S}(#1)}

\newcommand{\const}[2]{{\rm const}_{#2}{#1}}
\newcommand{\vect}[1]{\boldsymbol{#1}}
\newcommand{\vecT}[1]{({#1})}
\newcommand{\HNumber}[1]{H_{#1}}

\newcounter{linectr}

\newenvironment{myEnumerate}{\begin{list}{(\arabic{linectr})}{\usecounter{linectr}
\labelwidth1ex\itemsep0ex\labelsep1ex\leftmargin2ex\parskip0.0cm\topskip0cm\partopsep0cm
\listparindent0ex}}{\end{list}}

\title{Parameterized Telescoping Proves Algebraic Independence of Sums}

\author{Carsten Schneider}
\address{Research Institute for Symbolic Computation\\
J. Kepler University Linz\\
A-4040 Linz, Austria}
\email{Carsten.Schneider@risc.uni-linz.ac.at}
\thanks{Supported by the SFB-grant F1305 and the grant P20347-N18
of the Austrian FWF}



\begin{document}

\begin{abstract}
Usually creative telescoping is used to derive recurrences for sums.
In this article we show that the non-existence of a creative
telescoping solution, and more generally, of a parameterized
telescoping solution, proves algebraic independence of certain types
of sums. Combining this fact with summation-theory shows
transcendence of whole classes of sums.
Moreover, this result throws new light on the question why, e.g., Zeilberger's algorithm fails to find a recurrence with minimal order.
\end{abstract}

\maketitle

\section{Introduction}

Telescoping~\cite{Gosper:78} and creative
telescoping~\cite{Zeilberger:91,AequalB} for hypergeometric terms
and its
variations~\cite{PauleSchorn:95,Paule:95,PauleRiese:97,Bauer:99} are
standard tools in symbolic summation. All these techniques are
covered by the following formulation of the parameterized
telescoping problem: {\bf Given} sequences $f_1(k),\dots,f_d(k)$
over a certain field $\KK$, {\bf find}, if possible, constants
$c_1,\dots,c_d\in\KK$ and a sequence $g(k)$ such that
\begin{equation}\label{Equ:CreaTeleSequ}
g(k+1)-g(k)=c_1f_1(k)+\dots+c_df_d(k).
\end{equation}
If one succeeds in this task, one gets, with some mild extra
conditions, the sum-relation
\begin{equation}\label{Equ:GeneralRec}
g(n+1)-g(r)=c_1\ssum{k=r}nf_1(k)+\dots+c_d\ssum{k=r}nf_d(k)
\end{equation}
for some $r\in\NN=\{0,1,\dots\}$ big enough. Note that $d=1$ gives
telescoping. Moreover, given a bivariate sequence $f(m,k)$, one can
set $f_i(k):=f(m+i-1,k)$ which corresponds to creative telescoping.

Since Karr's summation algorithm~\cite{Karr:81} and its
extensions~\cite{Schneider:05a,Schneider:08a} can solve the
parameterized telescoping problem in the difference field setting of
\pisiSE-fields, we get a rather flexible algorithm which is
implemented in the package
\SigmaP~\cite{Schneider:04d,Schneider:07a}: the $f_i(k)$ can be
arbitrarily nested sums and products.

In this article we apply \pisiSE-field
theory~\cite{Karr:81,Schneider:T01} to get new theoretical insight:
If there is no solution to~\eqref{Equ:CreaTeleSequ} within a given
\pisiSE-field setting, then the sums in~\eqref{Equ:GeneralRec} can
be represented in a larger \pisiSE-field by transcendental
extensions; see Theorem~\ref{Thm:ConnectionDefSumAndProperSum}.
Motivated by this fact, we construct a difference ring monomorphism
which links elements from the larger \pisiSE-field to the sums
\begin{equation}\label{Equ:SumSet}
S_1(n)=\ssum{k=r}nf_1(k),\dots,S_d(n)=\ssum{k=r}nf_d(k)
\end{equation}
in the ring of sequences over $\KK$. In particular, this
construction transfers the transcendence properties from the
\pisiSE-world into the sequence domain. In order to accomplish this
task, we restrict to generalized d'Alembertian extensions, a
subclass of \pisiSE-extensions, which cover all those sum-product expressions that occurred in practical problem solving so far.

Summarizing, parameterized telescoping in combination with
\pisiSE-fields gives a criterion to check algorithmically the
transcendence of sums of type~\eqref{Equ:SumSet}; see
Theorem~\ref{Thm:MainSumTheorem}. Combining this criterion with
results from summation theory,
like~\cite{Abramov:71,Paule:95,Abramov:03b,Schneider:07d}, shows
that whole classes of sequences are transcendental. E.g., the
harmonic numbers $\set{H^{(i)}_n}{i\geq1}$ with
$H^{(i)}_n:=\sum_{k=1}^n\frac{1}{k^i}$ are algebraically independent
over $\QQ(n)$.

Moreover, we derive new insight for which sums creative telescoping, in particular Zeilberger's algorithm, finds the optimal recurrence and for which input classes it might fail to compute a recurrence with minimal order.

\smallskip

The general structure of this article is as follows. In
Section~\ref{Sec:PiSiDef} we present the basic notions of difference
fields, and we introduce \pisiSE-extensions together with the
subclass of generalized d'Alembertian extensions. In
Section~\ref{Sec:MainResults} we show the correspondence of
parameterized telescoping and the construction of a certain type of
\sigmaSE-extensions. In Section~\ref{Sec:HomoConstruction} we
construct a difference ring monomorphism that carries over the
transcendence properties from a given generalized d'Alembertian
extension to the ring of sequences. This leads to a transcendence
decision criterion of sequences in terms of generalized
d'Alembertian extensions in Section~\ref{Sec:MainApplication}. In Sections~\ref{Sec:Rat}--\ref{Sec:Nested} we
apply our criterion to the rational case, hypergeometric case, and to nested sums. Finally, we present the analogous criterion for products in Section~\ref{Sec:ProdTrans}.

\section{Basic notions: \pisiSE-extensions and generalized d'Alembertian
extension}\label{Sec:PiSiDef}

Subsequently, we introduce the basic concepts of difference fields
that shall pop up later.

A \notion{difference ring}\footnote{All fields and rings are of
characteristic $0$ and commutative} (resp. field)
$\dfield{\RR}{\sigma}$ is a ring $\RR$ (resp.~field) with a ring
automorphism (resp.~field automorphism) $\fct{\sigma}{\RR}{\RR}$.
The set of constants
$\const{\RR}{\sigma}=\set{k\in\RR}{\sigma(k)=k}$ forms a subring
(resp.\ subfield) of $\RR$. In this article we always assume that
$\const{\RR}{\sigma}$ is a field, which we usually denote by $\KK$.
We call $\const{\RR}{\sigma}$ the \notion{constant field} of
$\dfield{\RR}{\sigma}$.

A \notion{difference ring homomorphism} (resp.~difference ring
monomorphism) $\fct{\tau}{\RR_1}{\RR_2}$ between two difference
rings $\dfield{\RR_1}{\sigma_1}$ and $\dfield{\RR_2}{\sigma_2}$ is a
ring homomorphism (resp.~ring monomorphism) with the additional
property that $\tau(\sigma_1(f))=\sigma_2(\tau(f))$ for all
$f\in\RR_1$.

A difference ring (resp.~difference field) $\dfield{\EE}{\sigma}$ is
a \notion{difference ring extension} (resp.~\notion{difference field
extension}) of a difference ring (resp.~difference field)
$\dfield{\RR}{\sigma'}$ if $\RR$ is a subring (resp.~subfield) of
$\EE$ and $\sigma'(f)=\sigma(f)$ for all $f\in\RR$; since $\sigma$
and $\sigma'$ agree on $\RR$, we do not distinguish them anymore.

Now we are ready to define \pisiSE-extensions and generalized
d'Alembertian extensions in which we will represent our indefinite
nested sums and products.

\begin{Definition}
A difference field extension $\dfield{\FF(t)}{\sigma}$ of
$\dfield{\FF}{\sigma}$ is called a \notion{\pisiSE-extension} if
both difference fields share the same field of constants, $t$~is
transcendental over~$\FF$, and $\sigma(t)=t+a$ for some $a\in\FF^*$
(a sum) or $\sigma(t)=a\, t$ for some $a\in\FF^*$ (a product). If
$\sigma(t)/t\in\FF$ (resp. $\sigma(t)-t\in\FF$), we call the
extension also a \notion{\piE-extension}
(resp.~\notion{\sigmaSE-extension}). In short, we say that
$\dfield{\FF(t_1)\dots(t_e)}{\sigma}$ is a \notion\pisiSE-extension
(resp.~\piE-extension, \sigmaSE-extension) of $\dfield{\FF}{\sigma}$
if the extension is given by a tower of \pisiSE-extensions
(resp.~\piE-extensions, \sigmaSE-extensions). A
\notion{\pisiSE-field} $\dfield{\KK(t_1)\dots(t_e)}{\sigma}$ over
$\KK$ is a \pisiSE-extension of $\dfield{\KK}{\sigma}$ with constant
field $\KK$.
\end{Definition}

\begin{Example}\label{Exp:DField}
Consider the difference field $\dfield{\QQ(m)(k)(b)(h)}{\sigma}$
with $\sigma(k)=k+1$, $\sigma(b)=\frac{m-k}{k+1}b$,
$\sigma(h)=h+\frac{1}{k+1}$, and
$\const{\QQ(m)(k)(b)(h)}{\sigma}=\QQ(m)$. The extensions $k$, $b$,
and $h$ form \pisiSE-extensions over the fields below.
$\dfield{\QQ(m)(k)(b)(h)}{\sigma}$ is a \pisiSE-field over $\QQ(m)$.
\end{Example}

\noindent The following theorem tells us how one can check if an
extension is a \pisiSE-extension.

\begin{Theorem}[\cite{Karr:81}]\label{Thm:DecidePiSigmaExt}
Let $\dfield{\FF(t)}{\sigma}$ be a difference field extension of
$\dfield{\FF}{\sigma}$ with $\sigma(t)=\alpha\,t+\beta$
where $\alpha\in\FF^*$ and $\beta\in\FF$. Then:
\begin{myEnumerate}
\item This is a \sigmaSE-extension iff $\alpha=1$ and there is no $g\in\FF$ such that
$\sigma(g)-g=\beta$.

\item This is a \piE-extension iff $t\neq0$, $\beta=0$ and there are no $n\neq0$, $g\in\FF^*$ such that $\sigma(g)=\alpha^ng$.
\end{myEnumerate}
\end{Theorem}

\noindent The following remarks are in place:
\begin{enumerate}
\item If $\dfield{\FF}{\sigma}$ is a \pisiSE-field, algorithms are
available which make Theorem~\ref{Thm:DecidePiSigmaExt} completely
constructive; see~\cite{Karr:81,Schneider:05a}.

\item We emphasize that we have a first criterion for transcendence in
a difference field: if there is no telescoping solution, then we can
adjoin the sum as a transcendental extension without extending the
constant field. This criterion will be generalized to parameterized
telescoping; see Theorem~\ref{Thm:ConnectionDefSumAndProperSum}. For
the product case see Theorem~\ref{Thm:ConnectionProd}.
\end{enumerate}

Theorem~\ref{Thm:FirstOrderSumExt} states how solutions $g$ of
$\sigma(g)-g=f$ or $\sigma(g)=f g$ look like in certain types of
extensions. The first part follows by~\cite[Sec.~4.1]{Karr:81} and
the second part follows by~\cite[Lemma~6.8]{Schneider:05c}. These
results are crucial ingredients to prove
Theorems~\ref{Thm:ConnectionDefSumAndProperSum}
and~\ref{Thm:ConnectionProd}.

\begin{Theorem}\label{Thm:FirstOrderSumExt}
Let $\dfield{\FF(t_1,\dots,t_d)}{\sigma}$ be a \pisiSE-extension of
$\dfield{\FF}{\sigma}$ with constant field $\KK$ such that for all $1\leq i\leq d$ we have $\sigma(t_i)=\alpha_it_i+\beta_i$ with $\alpha_i,\beta_i\in\FF$.
Let $f\in\FF$ and $g\in\FF(t_1,\dots,t_d)$.
\begin{enumerate}
\item If $\sigma(g)-g=f$, then $g=\sum_{i=1}^dc_i\,t_i +w$ with
$w\in\FF$, $c_i\in\KK$; if $\alpha_i\neq1$, then $c_i=0$.

\item If $\sigma(g)=f\,g$, then $g=w\prod_{i=1}^dt_i^{c_i}$ with
$w\in\FF$ and $c_i\in\ZZ$; if $\beta_i\neq0$, then $c_i=0$.
\end{enumerate}
\end{Theorem}

%

Subsequently, we will restrict to the following type of extensions.

\begin{Definition}
We call
\pisiSE-extension $\dfield{\FF(t_1)\dots(t_e)}{\sigma}$ of
$\dfield{\FF}{\sigma}$ with $\sigma(t_i)=\alpha_i\,t_i+\beta_i$ \notion{generalized d'Alembertian extension} if
$\alpha_i\in\FF$ and $\beta_i\in\FF[t_1,\dots,t_{i-1}]$ for all
$1\leq i\leq e$.
\end{Definition}

\begin{Remark}\label{Rem:ReorderToForm}
Subsequently, we exploit the following fact: One can reorder a generalized d'Alembertian extension to
$\FF(p_1)\dots(p_u)(s_1)\dots(s_v)$ with $u,v\geq0$ where
$\frac{\sigma(p_i)}{p_i}\in\FF$ for $1\leq i\leq u$ and
$\sigma(s_i)-s_i\in\FF[p_1,\dots,p_u,s_1,\dots,s_{i-1}]$ for $1\leq
i\leq v$.
\end{Remark}

It is easy to see that $\dfield{\FF[t_1,\dots,t_e]}{\sigma}$ is a
difference ring extension of $\dfield{\FF}{\sigma}$. Moreover, if
$f\in\FF[t_1,\dots,t_e]$, then there are no solutions in
$\FF(t_1,\dots,t_e)\setminus\FF[t_1,\dots,t_e]$.

\begin{Theorem}\label{Thm:dAlembertProp}
Let $\dfield{\FF(t_1)\dots(t_e)}{\sigma}$ be a generalized
d'Alembertian extension of $\dfield{\FF}{\sigma}$ and suppose that
$g\in\FF(t_1)\dots(t_e)$. Then $\sigma(g)-g\in\FF[t_1,\dots,t_e]$ if
and only if $g\in\FF[t_1,\dots,t_e]$.
\end{Theorem}
\begin{proof}
The direction from left to right is clear by the definition of
generalized d'Alembertian extensions. We prove the other direction
by induction on the number of extensions. For $e=0$ nothing has to
be shown. Now suppose that the theorem holds for $e$ extensions and
consider the generalized d'Alembertian extension
$\dfield{\FF(t_1)\dots(t_{e+1})}{\sigma}$ of $\dfield{\FF}{\sigma}$.
By Remark~\ref{Rem:ReorderToForm} we can bring
$\FF(t_1)\dots(t_{e+1})$ to a form where all \sigmaSE-extensions are
on top. Write $t:=t_{e+1}$ and let $\sigma(t)=\alpha\,t+\beta$ with
$\alpha\in\FF^*$ and $\beta\in\FF[t_1,\dots,t_e]$. Now suppose that
$\sigma(g)-g=f$ where
$g\in\FF(t_1,\dots,t_e,t)\setminus\FF[t_1,\dots,t_e,t]$ and
$f\in\FF[t_1,\dots,t_e][t]$. Note that $g\in\FF(t_1,\dots,t_e)[t]$;
see, e.g.,~\cite[Lemma~3.1]{Schneider:07d}. Hence we can write
$g=\sum_{i=0}^d g_it^i$ with $g_i\in\FF(t_1,\dots,t_e)$. If $e=0$,
we are done. Otherwise, suppose that $e>0$ and let $j\geq0$ be
maximal such that
$g_j\in\FF(t_1,\dots,t_e)\setminus\FF[t_1,\dots,t_e]$. Define
$g':=\sum_{i=0}^{j}g_it^i\in\FF(t_1,\dots,t_e)[t]$ and
$f':=f-\big(\sigma(\sum_{i=j+1}^{d}g_it^i)-\sum_{i=j+1}^{d}g_it^i\big)\in\FF[t_1,\dots,t_e][t]$.
Since $\sigma(g')-g'=f',$  $\deg(f')\leq\deg(g')=j$. By coefficient
comparison we have
\begin{equation}\label{Equ:SubSpaceProb}
\alpha^j\sigma(g_j)-g_j=\phi\in\FF[t_1,\dots,t_{e}]
\end{equation}
where $\phi$ is the $j$th coefficient in $f'$. If $\alpha=1$ or
$j=0$, we can apply the induction assumption and conclude that
$g_j\in\FF[t_1,\dots,t_e]$, a contradiction. Otherwise, suppose that
$\alpha\neq1$ and $j\geq1$. Then by the assumption that all
\piE-extensions come first, it follows that $\sigma(t_i)/t_i\in\FF$
for all $1\leq i\leq e$. Reorder $\FF(t_1,\dots,t_e)$ such that
$g_j\notin\FF(t_1,\dots,t_{e-1})[t_e]$. By
Bronstein~\cite[Cor.~3]{Bron:00}, see
also~\cite[Cor.~1]{Schneider:04c}, we get $g_j=\frac{p}{t_e^m}$ for
some $m>0$ and\footnote{For a ring $\RR$ we define $\RR^*:=\RR\setminus\{0\}$.} $p\in\FF(t_1,\dots,t_{e-1})[t_e]^*$ with $t_e\nmid
p$. Hence
$\alpha^j\sigma(\frac{p}{t_e^m})-\frac{p}{t_e^m}=\frac{\alpha^j\sigma(p)-a^mp}{a^mt^m_e}=\phi$
with $a:=\frac{\sigma(t_e)}{t_e}\in\FF^*$. Since $t_e\nmid p$, also
$a\,t_e=\sigma(t_e)\nmid\sigma(p)$, and thus $t_e\nmid\sigma(p)$.
Since~\eqref{Equ:SubSpaceProb} and $m>0$, it follows that
$\alpha^j\sigma(p)-a^mp=0$, and hence
$\sigma(\frac{t_e^m}{p})=\alpha^j\,\frac{t_e^m}{p}$; a contradiction
to Theorem~\ref{Thm:DecidePiSigmaExt}.2 and the fact that
$\dfield{\FF(t_1,\dots,t_e)(t)}{\sigma}$ is a \piE-extension of
$\dfield{\FF(t_1,\dots,t_e)}{\sigma}$.
\end{proof}

Given a rational function field $\FF(t)$, we say that
$\frac{p}{q}\in\FF(t)$ is in reduced representation, if
$p,q\in\FF[t]$, $\gcd(p,q)=1$, and $q$ is monic.
The summation criterion
from~\cite{Abramov:71},\cite[Prop.~3.3]{Paule:95} and its
generalization to \pisiSE-extensions are substantial:

\begin{Theorem}(\cite[Cor.~5.1]{Schneider:07d})\label{Theorem:DispCrit}
Let $\dfield{\FF(t)}{\sigma}$ be a \pisiSE-extension of
$\dfield{\FF}{\sigma}$ and $\frac{p}{q}\in\FF(t)$ be in reduced
representation with $\deg(q)>0$ and with the property that either
$t\nmid q$ or $\frac{\sigma(t)}{t}\notin\FF$. If
$\gcd(\sigma^m(q),q)=1$ for all $m>0$, then there is no $g\in\FF(t)$
with $\sigma(g)-g=\frac{p}{q}$.
\end{Theorem}

Corollary~\ref{Cor:PiExtSum} is immediate; for a more general version see~\cite[Prop.~4.1.1]{Schneider:T01}.

\begin{Corollary}\label{Cor:PiExtSum}
Let $\dfield{\FF(t)}{\sigma}$ be a \piE-extension of $\dfield{\FF}{\sigma}$ with $\sigma(t)=\alpha\,t$, and let $w\in\FF$ and $g\in\FF(t)$. Then
$\sigma(g)-g=w\,t,$ iff $g=v\,t+c$ for $v\in\FF$, $c\in\const{\FF}{\sigma}$ where $\alpha\,\sigma(v)-v=w$.
\end{Corollary}
\begin{proof}
Suppose that $g\in\FF(t)$ with $\sigma(g)-g=w\,t$.
By Theorem~\ref{Theorem:DispCrit} $g\in\FF[t]$. Moreover, by~\cite[Cor.~2]{Karr:81}
or~\cite[Cor.~3]{Schneider:05b}, $g=v\,t+u$ with $v,u\in\FF$.
Plugging $g$ into $\sigma(g)-g=w\,t$ and doing coefficient
comparison shows that $u\in\const{\FF}{\sigma}$ and $\alpha\,\sigma(v)-v=w$. The other direction follows immediately.
\end{proof}

%

\section{Parameterized telescoping, \pisiSE-extensions and the ring
of sequences}\label{Sec:MainResults}

We get the following criterion to check transcendence in a given
difference field $\dfield{\FF}{\sigma}$.

\begin{Theorem}\label{Thm:ConnectionDefSumAndProperSum} Let
$\dfield{\FF}{\sigma}$ be a difference field  with constant field
$\KK$ and $\vecT{f_1,\dots,f_d}\in\FF^d$. The following statements
are equivalent.

\begin{enumerate}
\item There do not exist a $\vect{0}\neq\vecT{c_1,\dots,c_d}\in\KK^d$
and a $g\in\FF$ with
\begin{equation}\label{Equ:ParaDF}
\sigma(g)-g=c_1f_1+\dots+c_df_d.
\end{equation}

\item There is a \sigmaSE-extension
$\dfield{\FF(t_1)\dots(t_d)}{\sigma}$ of $\dfield{\FF}{\sigma}$ with
$\sigma(t_i)=t_i+f_i$ for $1\leq i\leq d$.

\end{enumerate}
\end{Theorem}
\begin{proof}
Suppose that~\eqref{Equ:ParaDF} holds for some
$\vect{0}\neq\vecT{c_1,\dots,c_d}\in\KK^d$ and $g\in\FF$. In
addition, assume that there exists a \sigmaSE-extension
$\dfield{\FF(t_1,\dots,t_d)}{\sigma}$ of $\dfield{\FF}{\sigma}$ with
$\sigma(t_i)=t_i+f_i$. Then
$\sigma(g)-g=\sum_{i=1}^dc_i\big(\sigma(t_i)-t_i\big)
=\sigma(\sum_{i=1}^d c_i\,t_i)-\sum_{i=1}^d c_i\,t_i$, and thus
$\sigma(\sum_{i=1}^d c_i\,t_i-g)=\sum_{i=1}^d c_i\,t_i-g.$ Since
$\const{\FF(t_1,\dots,t_d)}{\sigma}=\KK$, there is a $k\in\KK$ with
$\sum_{i=1}^d c_i\,t_i-g+k=0.$ Thus there are algebraic
relations in the $t_i$, a contradiction to the definition of \pisiSE-extensions.\\
Contrary, let $i\geq1$ be maximal such that
$\dfield{\FF(t_1,\dots,t_{i})}{\sigma}$ is a \sigmaSE-extension of
$\dfield{\FF}{\sigma}$; suppose that $i<d$. Then there is a
$g\in\FF(t_1,\dots,t_{i})$ with $\sigma(g)-g=f_{i+1}.$ By
Theorem~\ref{Thm:FirstOrderSumExt}.1 there are $c_j\in\KK$,
$h\in\FF$ with $g=h+\sum_{j=1}^{i}c_j\,t_j.$ This shows that
$\sigma(h)-h=f_{i+1}-\sum_{j=1}^{i}
c_j(\sigma(t_j)-t_j)=-c_1\,f_1-\cdots-c_{i}\,f_{i}+f_{i+1}$. Hence
we get a solution for~\eqref{Equ:ParaDF} with
$\vect{0}\neq\vecT{c_1,\dots,c_d}\in\KK^d$.
\end{proof}

Let $\KK$ be a field with characteristic zero. The set of all
sequences $\KK^{\NN}$ with elements $(a_n)_{n=0}^{\infty}=
(a_0,a_1,a_2,\dots)$, $a_i\in\KK$, forms a commutative ring by
component-wise addition and multiplication; the field $\KK$ can be
naturally embedded by identifying $k\in\KK$ with the sequence
$\vect{k}:=(k,k,k,\dots).$ In order to turn the
shift-operation
\begin{equation}\label{Equ:ShiftOp}
S:{(a_0,a_1,a_2,\dots)}\mapsto{(a_1,a_2,a_3,\dots)}
\end{equation}
to an automorphism, we follow the construction
from~\cite[Sec.~8.2]{AequalB}: We define an equivalence relation
$\sim$ on $\KK^{\NN}$ with $(a_n)_{n=0}^{\infty}\sim
(b_n)_{n=0}^{\infty}$ if there exists a $\delta\geq 0$ such that
$a_k=b_k$ for all $k\geq\delta$. The equivalence classes form a ring
which is denoted by $\seqR$; the elements of $\seqR$ will be
denoted, as above, by sequence notation. Now it is immediate that
$\fct{S}{\seqR}{\seqR}$ with~\eqref{Equ:ShiftOp} forms a ring
automorphism. The difference ring $\dfield{\seqR}{S}$ is called the
\notion{ring of $\KK$-sequences} or in short the \notion{ring of
sequences}.

\smallskip

The main construction of our article is that the polynomial ring
$\FF[t_1,\dots,t_e]$ of a generalized d'Alembertian extension
$\dfield{\FF(t_1)\dots(t_e)}{\sigma}$ of $\dfield{\FF}{\sigma}$ with
constant field $\KK$ can be embedded in the ring of sequences
$\dfield{\seqR}{S}$, provided that $\dfield{\FF}{\sigma}$ can be embedded in
$\dfield{\seqR}{S}$. More precisely, we will construct a difference ring
monomorphism $\fct{\tau}{\FF[t_1,\dots,t_e]}{\seqR}$ where the
constants $k\in\KK$ are mapped to $\vect{k}=(k,k,\dots)$. We will call such a difference ring homomorphism
(resp.~monomorphism) also a \notion{$\KK$-homomorphism}
(resp.~\notion{$\KK$-monomorphism}).
Then the main consequence is that the transcendence properties of generalized
d'Alembertian extensions, in particular
Theorem~\ref{Thm:ConnectionDefSumAndProperSum}, can be carried over
to $\seqR$; see Theorem~\ref{Thm:MainSumTheorem}.

\section{The monomorphism construction}\label{Sec:HomoConstruction}

In the following we will construct the $\KK$-monomorphism as
mentioned in the end of Section~\ref{Sec:MainResults}. Here we use
the following lemma which is inspired by \cite{PauleNemes:97}; the
proof is obvious.

\begin{Lemma}\label{Lemma:Link:SeqHom:Eval}
Let $\dfield{\RR}{\sigma}$ be a difference ring with constant field
$\KK$. If $\fct{\tau}{\RR}{\seqR}$ is a $\KK$-homomorphism, there is
a map $\fct{\ev}{\RR\times\NN}{\KK}$ with
\begin{equation}\label{Equ:EvDef}
\tau(f)=(\ev(f,0),\ev(f,1),\dots)
\end{equation}
for all $f\in\RR$ which has the following properties: For all
$c\in\KK$ there is a $\delta\geq0$ with
\begin{equation}\label{Ev:Const}
\forall i\geq\delta:\;\ev(c,i)=c;
\end{equation}
for all $f,g\in\RR$ there is a $\delta\geq 0$ with
\begin{align}
\forall i\geq\delta:\;\ev(f\,g,i)&=\ev(f,i)\,\ev(g,i),\label{Ev:Mult}\\
\forall i\geq\delta:\;\ev(f+g,i)&=\ev(f,i)+\ev(g,i)\label{Ev:Add};
\end{align}
and for all $f\in\RR$ and $j\in\ZZ$ there is a $\delta\geq 0$ with
\begin{equation}\label{Ev:Shift}
\forall i\geq\delta\;\ev(\sigma^j(f),i)=\ev(f,i+j).
\end{equation}
Conversely, if we have a map $\fct{\ev}{\RR\times\NN}{\KK}$
with~\eqref{Ev:Const},~\eqref{Ev:Mult}, \eqref{Ev:Add}
and~\eqref{Ev:Shift}, then the map $\fct{\tau}{\RR}{\seqR}$ defined
by~\eqref{Equ:EvDef} forms a $\KK$-homomorphism.
\end{Lemma}

\noindent In order to take into account the constructive aspects, we
introduce the following functions.

\begin{Definition}
Let $\dfield{\RR}{\sigma}$ be a difference ring and
$\fct{\tau}{\RR}{\seqR}$ be a $\KK$-homomor\-phism defined
by~\eqref{Equ:EvDef}. $\tau$ is called \notion{operation-bounded} by
$\fct{L}{\RR}{\NN}$ if for all $f\in\RR$ and $j\in\ZZ$ with
$\delta=\delta(f,j):=L(f)+\max(0,-j)$ we have~\eqref{Ev:Shift} and
for all $f,g\in\RR$ with $\delta=\delta(f,g):=\max(L(f),L(g))$ we
have~\eqref{Ev:Mult} and~\eqref{Ev:Add}. Moreover, we require that for all $f\in\RR$ and all $j\in\ZZ$ we have $L(\sigma^j(f))\leq L(f)+\max(0,-j)$; such a function is also
called \notion{o-function}. $\tau$ is called zero-bounded by
$\fct{Z}{\RR}{\NN}$ if for all $f\in\RR^*$ and all $i\geq Z(f)$ we
have $\ev(f,i)\neq0$; such a function is also called
\notion{z-function}.
\end{Definition}

\begin{Lemma}\label{Lemma:ZFuForField}
Let $\dfield{\RR}{\sigma}$ be a difference field with constant field
$\KK$. If $\fct{\tau}{\RR}{\seqR}$ is a $\KK$-homomorphism
with~\eqref{Equ:EvDef}, then for all $f\in\RR$ we have
$\ev(\tfrac{1}{f},i)=\frac{1}{\ev(f,i)}$ for big enough~$i$. In
particular, there is a z-function for~$\tau$.
\end{Lemma}
\begin{proof}
$\tau(f^{-1})$ is the inverse of $\tau(f)$, i.e.,
$\tau(\frac{1}{f})=\frac{1}{\tau(f)}$. Hence,
$\ev(\frac{1}{f},k)=\frac{1}{\ev(f,k)}$ for all $k\geq\delta$ for
some $\delta\geq0$. This implies $\ev(f,k)\neq0$ for all
$k\geq\delta$. Hence there is a z-function.
\end{proof}

The next lemma is the crucial tool to design step by step a
$\KK$-monomorphism for a generalized d'Alembertian extension. This
construction will be used in Theorem~\ref{Thm:MainSumTheorem}.

\begin{Lemma}\label{Lemma:LiftEvToPoly}
Let $\dfield{\FF(t_1)\dots(t_e)(t)}{\sigma}$ be a generalized
d'Alembertian extension of $\dfield{\FF}{\sigma}$ with
$\KK:=\const{\FF}{\sigma}$ and $\sigma(t)=\alpha\,t+\beta$. Let
$\fct{\tau}{\FF[t_1]\dots[t_e]}{S(\KK)}$ be a $\KK$-homomorphism with~\eqref{Equ:EvDef} together with an o-function $L$. Then:
\begin{myEnumerate}
\item There is a $\KK$-homomorphism  $\fct{\tau'}{\FF[t_1]\dots[t_e][t]}{S(\KK)}$ with
$\tau'(f)=(\ev'(f,l))_{l\geq0}$
for all $f\in\FF[t_1,\dots,t_e][t]$ such that $\tau'(f)=\tau(f)$ for all $f\in\FF[t_1,\dots,t_e]$; if $\beta=0$, $\ev'(t,k)\neq0$ for all $k\geq r$ for some $r\in\NN$. Such a $\tau'$ is uniquely determined by
\begin{equation}\label{Equ:SumProdHom}
\ev'(t,k)=\begin{cases}
\displaystyle c\,\prod_{i=r}^k\ev(\alpha,i-1)&\text{ if $\sigma(t)=\alpha\,t$}\\
\displaystyle\sum_{i=r}^k\ev(\beta,i-1)+c&\text{ if  $\sigma(t)=t+\beta$,}
\end{cases}
\end{equation}
up to the choice of $r\in\NN$ and $c\in\KK$; we require $c\neq0$, if $\beta=0$.

\item If $\tau$ is injective, $\tau'$ is injective.

\item There is an o-function for $\tau'$.

\item If there is a computable z-function for $\tau$ restricted on $\FF$ and a computable o-function for~$\tau$, then there is a computable
o-function for $\tau'$; $r$ in~\eqref{Equ:SumProdHom} can be computed.
\end{myEnumerate}
\end{Lemma}

\begin{proof}
{\bf(1)}~By Lemma~\ref{Lemma:ZFuForField} there is a z-function
$\fct{Z}{\FF}{\NN}$ for $\tau$ restricted on $\FF$. Let $\tau$ be
defined by~\eqref{Equ:EvDef}, denote $\RR:=\FF[t_1,\dots,t_e]$ and let $\sigma(t)=\alpha\,t+\beta$ with $\alpha\in\FF^*$ and
$\beta\in\RR$.\\
Let $\ev'(t,k)$ be defined as in~\eqref{Equ:SumProdHom} ($r$ will be specified later for the concrete cases $\alpha=1$ or $\beta=0$).
We extend $\ev$ from $\RR$ to $\RR[t]$ by
$$\ev'(\sum_{i=0}^nf_it^i,k)=\sum_{i=0}^n\ev(f_i,k)\ev'(t,k)^i.$$
First suppose that $\alpha=1$. Let $r:=L(f)+1$ and
consider the sequence given by~\eqref{Equ:SumProdHom} for some $c\in\KK$. Let $j\geq0$. Then by construction and by the choice of $r$ we have for all $k\geq r$:
\begin{align*}
\ev'(\sigma^j(t),k)&=\ev'(t+\ssum{i=0}{j-1}\sigma^i(\beta),k)=\ev'(t,k)+
\ssum{i=0}{j-1}\ev(\sigma^i(\beta),k)+c\\
&=\ev'(t,k)+\ssum{i=0}{j-1}\ev(\beta,k+i)+c=\ev'(t,k+j)=\ev'(t,k+j).
\end{align*}
\noindent Similarly, if
$j<0$, then $\ev'(t,k+j)=\ev'(\sigma^j(t),k)$ for all $k\geq r-j$.
This proves~\eqref{Ev:Shift} for $f=t$ and all $j\in\ZZ$ with
$\delta=r+\max(-j,0)$ ($\ev$ is replaced by $\ev'$). Now suppose that $\beta=0$. Let
$r:=\max(Z(\alpha),L(f))+1$ and $c\in\KK^*$, and consider the sequence given by~\eqref{Equ:SumProdHom}.
Analogously, it follows that for all $j\in\ZZ$ we have
$\ev'(t,k+j)=\ev'(\sigma^j(t),k)$ for all $k\geq r+\max(-j,0)$.
Moreover, since $\ev(\alpha,i-1)\neq0$ for
all $i\geq r$, $\ev'(t,k)\neq0$ for all $k\geq r$.
Moreover, if we choose $\delta\geq r$ big enough (depending on the
$f_i$), we get~\eqref{Ev:Shift} for $f=\sum_{i=0}^nf_it^i$.
Similarly, we can find for all $f,g\in\RR[t]$ a $\delta\geq0$
with~\eqref{Ev:Mult} and~\eqref{Ev:Add}. Moreover,~\eqref{Ev:Const}
holds, since $\ev'$ restricted on $\RR$ equals $\ev$. Summarizing,
if we define $\fct{\tau'}{\RR[t]}{\seqR}$ by $\tau'(f)=(\ev'(f,i))_{i\geq0}$
for all $f\in\RR[t]$, $\tau'$ forms a $\KK$-homomorphism by
Lemma~\ref{Lemma:Link:SeqHom:Eval}. Note: if $\beta=0$, then $\ev'(t,k)\neq0$ for all $k\geq r$.\\
Furthermore, this construction is unique up to $c$ and $r$ in~\eqref{Equ:SumProdHom}. Namely, take any other $\tau_2$ with $\tau_2(f)=\tau(f)$ for $f\in\FF[t_1,\dots,t_e]$ and define $T:=\tau_2(t)$; if $\beta=0$, we require in addition that $T$ is nonzero from a certain point on. Then $S(T)=S(\tau_2(t))=\tau_2(\sigma(t))=\tau_2(\alpha\,t+\beta)=\tau_2(\alpha)T+\tau(\beta)$.
Note that $\tau_2(1)=\vect{1}$ and $\tau_2(0)=\vect{0}$. Hence, if $\alpha=1$, $S(T)=T+\tau(\beta)$, and therefore $S(T-\tau(t))=T-\tau(t)$, i.e., $T=\tau(t)+\vect{d}$ for some constant $d\in\KK$. Similarly, if $\beta=0$, $S(T)=\tau(\alpha)T$. Since $\tau'(t)$ is non-zero from the point $r$ on, one can take the inverse $1/\tau'(t)$ and gets $S(\frac{T}{\tau'(t)})=\frac{T}{\tau'(t)}$. Hence $\frac{T}{\tau'(t)}=\vect{d}$ with $d\in\KK$, i.e., $T=\vect{d}\tau'(t)$. Since $T$ is nonzero for almost all entries, $d\neq0$. This shows that $\tau_2$ can be defined by~\eqref{Equ:SumProdHom} up to a constant $d\in\KK$; $d\neq0$, if $\beta=0$. Note: a different $r$ can be compensated by an appropriate choice of $d$.\\
{\bf(2)}~Suppose that $\tau$ is a $\KK$-monomorphism, but the
extended $\KK$-homomor\-phism $\tau'$ is not injective. Then take
$f=\sum_{i=0}^nf_it^i\in\RR[t]^*$ with $\tau'(f)=\vect{0}$ where $\deg(f)=n$ is minimal. Note that $f\notin\RR$, otherwise $\tau'(f)=\tau(f)=\vect{0}$; a contradiction that $\tau$ is injective. Hence, $n>0$. Define
\begin{equation}\label{Equ:HSegRed}
h:=\sigma(f_n)\alpha^n\,f-f_n\sigma(f)=\sigma(f_n)\alpha^n\ssum{i=0}{n}f_it^i-f_n\ssum{i=0}{n}\sigma(f_i)(\alpha\,t+\beta)^i\in\RR[t].
\end{equation}
Since $\vect{0}=S(\vect{0})=S(\tau'(f))=\tau'(\sigma(f))$, we have
$\tau'(h)=\tau(\sigma(f_n)\alpha^n)\tau'(f)-\tau(f_n)\tau'(\sigma(f))=\vect{0}$.
In addition, $\deg(h)<n$ by construction. It follows that  $h=0$, otherwise we get a contradiction to the minimality of $n$. With~\eqref{Equ:HSegRed} we get $\sigma(f)/f\in\FF(t_1)\dots(t_e)$
with $f\notin\FF(t_1)\dots(t_e)$. If $t$ is a \sigmaSE-extension, we
get a contradiction by Theorem~\ref{Thm:FirstOrderSumExt}.1.
Otherwise, suppose that $t$ is a \piE-extension. W.l.o.g.\ suppose that all \piE-extensions come first, say $t_1,\dots,t_r$ ($r\geq0$). Then $f=u\,t_1^{m_1}\dots t_r^{m_r}t^n$
with $m_1\,\dots,m_r\in\NN$ and $u\in\FF^*$ by Theorem~\ref{Thm:FirstOrderSumExt}.2. Hence
$\vect{0}=\tau'(f)=\tau(u)\,\tau(t_1)^{m_1}\dots\tau(t_r)^{m_r}
\tau'(t)^n$. Note that for $1\leq i\leq r$, $\tau(t_i)$ is non-zero from a point on (for all $r\geq0$, $S^r(\tau(t_i))=\vect{v}\tau(t_i)$ for some $\vect{v}\in\seqR$; hence if infinitely many zeros occur, we can variate $r$ to prove $\tau(t_i)=\vect{0}$; a contradiction). Since also $\tau(u)\neq\vect{0}$
($\tau$ is injective and $u\neq0$), $\tau'(t)$ has infinitely many zeros, a contradiction to our construction of $\tau'$. Summarizing, $\tau'$ is injective.\\
\noindent{\bf(3)}~Note that the $r$ for~\eqref{Equ:SumProdHom} can be defined by
$r:=\max(Z(\alpha),L(\alpha))+1$ or $r:=L(\beta)+1$, respectively.
Define $\fct{L'}{\RR[t]}{\NN}$ by
$$L'(f)=\begin{cases}
L(f)&\text{if }f\in\RR,\\
\max(r,L(f_1),\dots,L(f_n))&\text{if
$f=\sum_{i=0}^nf_it^i\notin\RR$}.
\end{cases}$$
Then one can check that $L'$ is an o-function for $\tau'$. E.g., for
$f=\sum_{i=0}^mf_it^i, g=\sum_{i=0}^ng_it^i$:
\begin{align*}
\ev'(f\,g,k)&=
\ev'(\ssum{j=0}{m+n}t^j\ssum{i=0}{j}f_ig_{j-i},k)=\ssum{j=0}{m+n}\ev'(t,k)^j\ssum{i=0}{j}\ev(f_ig_{j-i},k)\\
&=
\Big(\ssum{i=0}{m}\ev(f_i,k)\ev'(t,k)^i\Big)\Big(\ssum{j=0}{n}\ev(g_j,k)\ev'(t,k)^j\Big)=
\ev(f,k)\ev(g,k)
\end{align*}
for all
$k\geq\max(r,L(f_0),\dots,L(f_m),L(g_0),\dots,L(g_n))=\max(L'(f),L'(g))$.\\
{\bf(4)}~In particular, if $L$ and $Z$ are computable, then $L'$ is computable; by construction an appropriate $r$ in~\eqref{Equ:SumProdHom} can be computed.
\end{proof}

By iterative application of the previous lemma we arrive at

\begin{Theorem}\label{Thm:EmbedTheorem}
Let $\dfield{\FF(t_1)\dots(t_e)}{\sigma}$ be a generalized
d'Alembertian-extension of $\dfield{\FF}{\sigma}$ with
$\KK:=\const{\FF}{\sigma}$.
If there is a $\KK$-homomorphism/$\KK$-monomorphism $\fct{\tau}{\FF}{\seqR}$ with an o-function $L$, then there is a
$\KK$-homomorphism/$\KK$-monomorphism $\fct{\tau'}{\FF[t_1,\dots,t_e]}{\seqR}$ with
$\tau'(f)=\tau(f)$ for all $f\in\FF$ together with an o-function $L'$ for $\tau'$. If $L$ is computable and $\tau$ has a
computable z-function, $L'$ is computable.
\end{Theorem}

In order to apply Theorem~\ref{Thm:EmbedTheorem}, we must be able to embed the ground field $\dfield{\FF}{\sigma}$ with $\KK=\const{\FF}{\sigma}$ in the ring of sequences $\dfield{\seqR}{S}$ by a
$\KK$-monomorphism. We give a criterion when this is possible in
Theorem~\ref{Thm:LiftEvToRat}. Applying this result we get, e.g.,
$\KK$-monomorphisms for the rational case, the $q$-rational case and
the mixed case.

\begin{Theorem}\label{Thm:LiftEvToRat}
Let $\dfield{\FF(t)}{\sigma}$ be a \pisiSE-extension of
$\dfield{\FF}{\sigma}$ with constant field $\KK$ and let
$\fct{\tau}{\FF[t]}{\seqR}$ be a $\KK$-homomorphism (resp.
$\KK$-monomorphism).
\begin{enumerate}
\item There is a z-function for $\tau$ if and only if there is a
$\KK$-homomorphism (resp.~$\KK$-monomorphism)
$\fct{\tau'}{\FF(t)}{\seqR}$.

\item Let $Z$ and $L$ be z- and o-functions for $\tau$. Then there is
a $\KK$-homomorphism (resp.~$\KK$-monomorphism)
$\fct{\tau'}{\FF(t)}{\seqR}$ with a $z$-function $Z'$ and an
$o$-function $L'$. If $Z$ and $L$ are computable, then $Z'$ and $L'$
are computable.
\end{enumerate}
\end{Theorem}
\begin{proof}
(1)~The direction from right to left follows by
Lemma~\ref{Lemma:ZFuForField}. Suppose that $Z$ is a z-function
for~$\tau$. Let $\frac{p}{q}\in\FF(t)$ be in reduced representation.
Then we extend $\ev$ to $\FF(t)$ by
\begin{equation}\label{Equ:RatDef}
\ev(\frac{p}{q},k)=\begin{cases}
0 &\text{if $k<Z(q)$}\\
\frac{\ev(p,k)}{\ev(q,k)}&\text{if $k\geq Z(q)$}.
\end{cases}
\end{equation}
The
properties~\eqref{Ev:Const},~\eqref{Ev:Mult},~\eqref{Ev:Add},~\eqref{Ev:Shift}
can be carried over from $\FF[t]$ to $\FF(t)$. By
Lemma~\ref{Lemma:Link:SeqHom:Eval} we get a $\KK$-homomorphism
$\fct{\tau'}{\FF(t)}{\seqR}$ with~\eqref{Equ:EvDef}. Finally,
suppose that $\tau$ is injective. Take $f=\frac{p}{q}$ in reduced
form such that $\vect{0}=\tau'(f)=\frac{\tau(p)}{\tau(q)}$. Since
$\ev(q,k)\neq0$ for all $k\geq Z(q)$, $\tau(p)=0$. As $\tau$ is
injective, $p=0$ and thus $f=0$. This proves that
$\tau'$ is injective.\\
(2)~Let $L$ and $Z$ be o- and $z$-functions for $\tau$,
respectively. Then we extend them to $\FF(t)$ by
\begin{equation}\label{Equ:LZToRat}
\begin{split}
Z'(\frac{p}{q})&=
\begin{cases}
Z(p)&\text{if $q=1$}\\
\max(Z(p),Z(q))&\text{if $q\neq1$}, \end{cases}\\
L'(\frac{p}{q})&=\begin{cases}L(p)&\text{if
$q=1$}\\
\max(L(p),L(q),Z(q))&\text{if $q\neq1$}\end{cases}
\end{split}
\end{equation}
where $\frac{p}{q}\in\FF(t)$ is in reduced representation. By
construction $Z'$ and $L'$ are $z$- and $o$-functions for $\tau'$.
If $L$ and $Z$ are computable, then $Z'$ and $L'$ are computable.
\end{proof}

\begin{Remark}
Given the $\KK$-homomorphism (resp.~$\KK$-monomorphism) $\fct{\tau}{\FF[t]}{\seqR}$, the $\KK$-homomorphism (resp.~$\KK$-monomorphism) $\fct{\tau'}{\FF(t)}{\seqR}$ with $\tau'(p)=\tau(p)$ for all $p\in\FF[t]$ is uniquely determined by~\eqref{Equ:RatDef} -- up to the choice of the z-function $Z$.
\end{Remark}

\begin{Example}\label{Exp:RationalCase}
Let $\dfield{\KK(n)}{\sigma}$ be the \pisiSE-field over $\KK$ with
$\sigma(n)=n+1$. Then by our construction we get a
$\KK$-monomorphism $\fct{\tau}{\KK(n)}{\seqR}$ with computable o-
and z-functions as follows: Start with the $\KK$-monomorphism
$\fct{\tau}{\KK}{\seqR}$ with $\tau(k)=\vect{k}=(k,k,\dots)$ and take the o-function $L(k)=0$ and the
z-function $Z(k)=0$ for all $k\in\KK$. By
Lemma~\ref{Lemma:LiftEvToPoly} we get the $\KK$-monomorphism
$\fct{\tau}{\KK[n]}{\seqR}$ defined by $\ev(p,k)=p(k)$ for all
$p\in\KK[n]$ and all $k\geq0$. The resulting o-function is $L(p)=0$
for all $p\in\KK[n]$. Note that the $z$-function exists since
$p(n)\in\KK[n]$ can have only finitely many roots. The nonnegative
integer roots can be easily computed; see,
e.g.,~\cite[page~80]{AequalB}. Hence by
Theorem~\ref{Thm:LiftEvToRat} we can lift the $\KK$-monomorphism
from $\KK[n]$ to $\KK(n)$ with the o-function $L'$  and
z-function $Z'$ given by~\eqref{Equ:LZToRat}.
\end{Example}

\begin{Lemma}\label{Lemma:mixedInduction}
Let $\dfield{\KK(q)(t_1)\dots(t_e)(t)}{\sigma}$ be a \pisiSE-field
over the rational function field $\KK(q)$ where
$\sigma(t_i)=\alpha_it_i+\beta_i$ with $\alpha_i,\beta_i\in\KK$ and
$\sigma(t)=q\,t$. If there is a $\KK$-monomorphism
$\fct{\tau}{\KK(q)(t_1)\dots(t_e)}{\seqP{\KK(q)}}$ with (computable)
o- and z-functions, there is a $\KK$-monomor\-phism
$\fct{\tau}{\KK(q)(t_1)\dots(t_e)(t)}{\seqP{\KK(q)}}$ with
(computable) o- and z-functions.
\end{Lemma}
\begin{proof}
By Lemma~\ref{Lemma:LiftEvToPoly} there is a $\KK$-monomorphism
$\fct{\tau'}{\KK(q)(t_1)\dots(t_e)[t]}{\seqP{\KK(q)}}$ with an
o-function $L'$; $L'$ is computable if $L$ is computable. In this
construction we can take $\ev(t,k)=q^k$.
By~\cite[Sec.~3.7]{Bauer:99} there is a $Z'$-function for
$\KK(q)(t_1)\dots(t_e)[t]$; it is computable, if $Z$ is computable.
By Theorem~\ref{Thm:LiftEvToRat} we get a $\KK$-monomorphism from
$\KK(q)(t_1)\dots(t_e)(t)$ to $\seqP{\KK(q)}$ with o- and
z-functions; they are computable, if $L',Z'$ are computable.
\end{proof}

\noindent By Expample~\ref{Exp:RationalCase} and iterative
application of Lemma~\ref{Lemma:mixedInduction} based
on~\cite{Bauer:99} we get the mixed case.

\begin{Corollary}\label{Cor:MixedCase}
Let $\dfield{\KK(n)(t_1)\dots(t_e)}{\sigma}$ be a \pisiSE-field over
a rational function field $\KK:=\KK'(q_1)\dots(q_e)$ where
$\sigma(n)=n+1$ and $\sigma(t_i)=q_it_i$ for all $1\leq i\leq e$.
Then there is a $\KK$-monomorphism
$\fct{\tau}{\KK(n)(t_1)\dots(t_e)}{\seqP{\KK}}$ with a computable
o-function and z-function.
\end{Corollary}

\noindent Note that the use of asymptotic arguments might produce
$\KK$-monomorphisms for more general \pisiSE-fields. An open
question is, if any \pisiSE-field over $\KK$ can be embedded in
$\seqR$.

\section{A criterion to check  algebraic independence}\label{Sec:MainApplication}

We consider the following application.
Given sums of the type~\eqref{Equ:SumSet}, we start with an appropriate \pisiSE-field
$\dfield{\FF}{\sigma}$ (e.g., the rational case, $q$-rational case, or the mixed case) and construct, if possible, a generalized d'Alembertian
extension $\dfield{\FF(t_1)\dots(t_e)}{\sigma}$ of
$\dfield{\FF}{\sigma}$ together with a $\KK$-monomorphism such that for each $1\leq i\leq d$:
$f_i\in\FF[t_1,\dots,t_e]$ with $\ev(f_i,k)=f_i(k)$ for all $k\geq r$ for some $r\in\NN$. Here one
must choose within the monomorphism construction the initial values
$c$ in~\eqref{Equ:SumProdHom} accordingly.

\medskip

\noindent{\it Remark.} In~\SigmaP\ this translation mechanism is done automatically; see, e.g., Section~\ref{Sec:Nested}.

\medskip

Then one can prove or disprove the transcendence of the sums~\eqref{Equ:SumSet} by
showing the non-existence or existence of a parameterized
telescoping solution~\eqref{Equ:ParaDF} in $\FF[t_1,\dots,t_e]$. This fact can be summarized in

\begin{Theorem}[{\bf Main result}]\label{Thm:MainSumTheorem}
Let $\dfield{\FF(t_1)\dots(t_e)}{\sigma}$ be a generalized
d'Alembertian-extension of $\dfield{\FF}{\sigma}$ with
$\KK:=\const{\FF}{\sigma}$, and let
$\fct{\tau}{\FF[t_1,\dots,t_e]}{\seqR}$ be a $\KK$-monomorphism
with~\eqref{Equ:EvDef} together with an o-function; let
$\vecT{f_1,\dots,f_d}\in\FF[t_1,\dots,t_e]^d$.  Then the following
statements are equivalent:
\begin{enumerate}
\item There are no $g\in\FF[t_1,\dots,t_e]$ and $\vect{0}\neq\vecT{c_1,\dots,c_d}\in\KK^d$
with~\eqref{Equ:ParaDF}.

\item The sequences $\{(S_1(n))_{n\geq0},\dots,(S_d(n))_{n\geq0}\}$ given by
\begin{equation}\label{Equ:SumEvEmbed}
S_1(n):=\sum_{k=r}^n\ev(f_1,k),\dots,S_d(n):=\sum_{k=r}^n\ev(f_d,k).
\end{equation}
for some $r$ big enough, are algebraically independent over $\tau(\FF[t_1,\dots,t_e])$.
\end{enumerate}

\end{Theorem}

\begin{proof}
Suppose~(1) holds. Then by Theorem~\ref{Thm:dAlembertProp} and Theorem~\ref{Thm:ConnectionDefSumAndProperSum} there is
the \sigmaSE-extension $\dfield{\FF(t_1)\dots(t_e)(s_1)\dots(s_d)}{\sigma}$ of $\dfield{\FF(t_1)\dots(t_e)}{\sigma}$ with
\begin{equation*}
\sigma(s_1)=s_1+f_1,\dots,\sigma(s_d)=s_d+f_d.
\end{equation*}
Moreover, there is a $\KK$-monomorphism $\fct{\tau'}{\FF[t_1,\dots,t_e][s_1,\dots,s_d]}{\seqR}$
defined by $\tau'(f)=\tau(f)$ for all $f\in\FF[t_1,\dots,t_e]$ and $\tau'(s_i)=(S_i(n))_{n\geq0}$ for $1\leq i\leq d$ where~\eqref{Equ:SumEvEmbed}
for some $r$ big enough. Since
$R[s_1,\dots,s_d]$ is a polynomial ring over $R:=\FF[t_1,\dots,t_e]$, (2) follows by the $\KK$-monomorphism $\tau'$.\\
Conversely,
suppose that~(1) does not hold. Then we
get~\eqref{Equ:CreaTeleSequ} with $g(k):=\ev(g,k)$ and
$f_i(k):=\ev(f_i,k)$ with $k$ big enough, say $k\geq r$. Summing
this equation over $r\leq k\leq n$ gives a relation of the
form~\eqref{Equ:GeneralRec}, i.e., the sums in~\eqref{Equ:SumSet}
are algebraic. Thus~(2) does not hold.
\end{proof}

From the algorithmic point of view, \SigmaP\ can check the non-existence of a solution of~\eqref{Equ:ParaDF}, which then implies the transcendence of the sums~\eqref{Equ:SumEvEmbed}. Note that an appropriate $r$ in~\eqref{Equ:SumEvEmbed} is computable if $\tau$ has a computable o- and z-function.

Besides this, restricting the $f_i$ to elements with certain
structure allows to predict the non-existence of a solution of~\eqref{Equ:ParaDF}. In this way, we are able to classify various types of sums to be algebraically independent.
Subsequently, we will explore these aspects for various types of sums.

\section{Rational sums}\label{Sec:Rat}
Applying Theorem~\ref{Thm:MainSumTheorem} together with
Example~\ref{Exp:RationalCase} gives the following theorem.
\begin{Theorem}\label{Thm:RatCase}
Let $f_1(k),\dots,f_d(k)\in\KK(k)$. If there are no $g(k)\in\KK(k)$
and $c_1,\dots,c_d\in\KK$ with~\eqref{Equ:CreaTeleSequ} then the
sequences~\eqref{Equ:SumSet}, for some $r$ big enough, are
algebraically independent over $\KK(n)$, i.e., there is no
polynomial $P(x_1,\dots,x_d)\in\KK(n)[x_1,\dots,x_d]^*$ with
\begin{equation}\label{Equ:PolyRel}
P(S_1(n),\dots,S_d(n))=0\;\;\;\forall n\geq0.
\end{equation}
\end{Theorem}

\begin{Corollary}\label{Cor:RatTrans}
Let $p_1(k),p_2(k),\ldots\in\KK[k]^*$,
$u_1(k),u_2(k),\ldots\in\KK[k]^*$ and $q\in\KK[k]^*$ with
$\deg(q)>0$ and $\gcd(p_i,q)=\gcd(u_i,q)=1$ for all $i\geq1$;
suppose that $q(r)\neq0$ for all $r\in\NN^*$ and
$\gcd(q(k),q(k+r))=1$ for all $r\in\NN^*$. Then the sums
$$S_1(n):=\sum_{k=1}^nu_1(k)\left(\frac{p_1(k)}{q(k)}\right),
S_2(n):=\sum_{k=1}^nu_2(k)\left(\frac{p_2(k)}{q(k)}\right)^2,\dots$$
are algebraically independent over $\KK(n)$, i.e., there is no
$P(x_1,\dots,x_d)\in\KK(n)[x_1,\dots,x_d]^*$ for some $d\geq1$
with~\eqref{Equ:PolyRel}.
\end{Corollary}

\begin{proof}
Denote $f_i(k):=u_i\big(\frac{p_i}{q}\big)^i$ and suppose there are
$g(k)\in\KK(k)$ and $c_i\in\KK$ with~\eqref{Equ:CreaTeleSequ} where
$d\geq1$ is minimal. Then it follows that
$$g(k+1)-g(k)=\frac{c_1u_1p_1q^{d-1}+c_2u_2p_2^2q^{d-2}+\dots+c_du_dp_d^d}{q^d}=:\frac{v}{q^d}.$$
Since $c_d\neq0$, $\gcd(q,c_du_dp_d^d)=1$. Hence $\gcd(v,q)=1$, and
thus $\gcd(v,q^d)=1$. By Theorem~\ref{Theorem:DispCrit}, $g(k)\in\KK(k)$ cannot exist; a contradiction. The
corollary follows by Theorem~\ref{Thm:RatCase}.
\end{proof}

\begin{Example}
Choosing $p_i=u_i=1, q=k$ in Corollary~\ref{Cor:RatTrans} proves
that the generalized harmonic numbers $\{H^{(i)}_n|i\geq1\}$ are
algebraically independent over $\KK(n)$.
\end{Example}

\noindent Applying Theorem~\ref{Thm:MainSumTheorem} together with
Corollary~\ref{Cor:MixedCase} accordingly produces the $q$-versions
and the mixed versions of Theorem~\ref{Thm:RatCase} and
Corollary~\ref{Cor:RatTrans}. A typical application is
Example~\ref{Exp:qHarmonic}.

\begin{Example}\label{Exp:qHarmonic}
The $q$-harmonic numbers
$\{\sum_{k=1}^n\frac{1}{(1-q^k)^i}|i\geq1\}$ (or for instance the variations
$\{\sum_{k=1}^n\big(\frac{q^{k}}{1-q^k}\big)^i|i\geq1\}$) are
algebraically independent over $\KK(q^k)$.
\end{Example}

\noindent Completely analogously to Corollary~\ref{Cor:RatTrans} one
can show the following corollary

\begin{Corollary}
Let $p_1(k),q_1(k),\dots,p_d(k),q_d(k)\in\KK[k]^*$ with
$\deg(q_i)>0$ and $\gcd(p_i,q_i)=1$. Suppose that $q_i(k)\neq0$ for
all $k\in\NN$ and that $\gcd(q_i(k+r),q_j(k))=1$ for all $r\in\ZZ$
and all $1\leq i<j\leq d$. Then the sums
$$S_1(n):=\sum_{k=1}^n\frac{p_1(k)}{q_1(k)},\dots,S_d(n):=\sum_{k=1}^n\frac{p_d(k)}{q_d(k)}$$
are algebraically independent over $\KK(n)$, i.e., there is no
$P(x_1,\dots,x_d)\in\KK(n)[x_1,\dots,x_d]^*$
with~\eqref{Equ:PolyRel}.
\end{Corollary}


\section{Hypergeometric sums and the minimality of recurrences}\label{Sec:Hyp}

Suppose that $f(k)$ is a hypergeometric term in $k$, i.e., there is
an $\alpha\in\KK(k)$ with $\alpha(r):=\frac{f(r+1)}{f(r)}$ for all
$r$ big enough; in short we also write
$\alpha(k):=\frac{f(k+1)}{f(k)}$ to define the rational function
$\alpha\in\KK(k)$. In this context the following result is important. $f(k)$ can be represented with $t$ in the
\pisiSE-field $\dfield{\KK(k)(t)}{\sigma}$ over $\KK$ with
$\sigma(k)=k+1$ and $\sigma(t)=\alpha\,t$ if and only if there are
no $r(k)\in\KK(k)$ and no root of unity $\gamma$ with $f(k)=\gamma^k
r(k)$; see~\cite[Thm.~5.4]{Schneider:05c}. In the following, we exclude this special case.

Subsequently, we exploit Corollary~\ref{Cor:PiExtSum} for the hypergeometric case.

\begin{Corollary}\label{Cor:HypSum}
Let $f(k)$ be a hypergeometric term such that $f(k)\neq \gamma^kr(k)$ for
all $r(k)\in\KK(k)$ and all roots of unity $\gamma$, and consider the \pisiSE-field
$\dfield{\KK(k)(t)}{\sigma}$ over $\KK$ with $\sigma(k)=k+1$ and
$\sigma(t)=\frac{f(k+1)}{f(k)}t$.
Then
$\sigma(g)-g=w\,t,$ if and only if $g=v\,t+c$ for $v\in\KK(k)$, $c\in\KK$ where
\begin{equation}\label{GosperEqu}
\alpha(k)\,v(k+1)-v(k)=w.
\end{equation}
\end{Corollary}

We note that~\eqref{GosperEqu} is nothing else than the basic ansatz~\cite[Equ.~5.2.2]{AequalB} of Gosper's algorithm. In a nutshell, Gosper's algorithm (and also Zeilberger's algorithm) check the existence of a solution in the corresponding \pisiSE-field
$\dfield{\KK(k)(t)}{\sigma}$.

As a consequence, Theorem~\ref{Thm:MainSumTheorem} can be simplified to the following version.

\begin{Theorem}
Let $f(k)$ be a hypergeometric term such that $f(k)\neq \gamma^kr(k)$ for
all $r(k)\in\KK(k)$ and all roots of unity $\gamma$, and consider the \pisiSE-field
$\dfield{\KK(k)(t)}{\sigma}$ over $\KK$ with $\sigma(k)=k+1$ and
$\sigma(t)=\frac{f(k+1)}{f(k)}t$. Let $r_i(k)\in\KK(k)$ for $1\leq
i\leq d$ and set $f_i:=r_it\in\KK(k)(t)$. If there are no
$c_i\in\KK$ and $w\in\KK(k)$ with $g:=w\,t$ such
that~\eqref{Equ:ParaDF}, then the following sequences, for $r$ big
enough, are algebraically independent over $\KK(n)$:
$$f(n),\;
S_1(n)=\sum_{k=r}^nr_1(k)f(k),\dots,S_d(n)=\sum_{k=r}^br_d(k)f(k),$$
i.e., there is no
$P(x_0,x_1,\dots,x_d)\in\KK(n)[x_0,x_1,\dots,x_d]^*$ with
\begin{equation}\label{Equ:HypRel}
P(f(n),S_1(n),\dots,S_d(n))=0\;\;\;\forall n\geq0.
\end{equation}
\end{Theorem}
\begin{proof}
Suppose there is a solution $c_i\in\KK$ and $g\in\KK(k)(t)$
with~\eqref{Equ:ParaDF}. By~Corollary~\ref{Cor:PiExtSum},
$\sigma(w t)-wt=t\sum_{i=1}^dc_i
r_i=\sum_{i=1}^dc_i\,f_i$ for some $w\in\FF$; a contradiction to the assumption.
Applying Theorem~\ref{Thm:MainSumTheorem} and choosing an
appropriate $\KK$-monomorphism proves the theorem.
\end{proof}

In particular, in the context of finding recurrences we obtain the following result.

\begin{Corollary}\label{Cor:Contiguous}
Let $f(\vect{m},k)$ be a hypergeometric term in
$\vect{m}=\vecT{m_1,\dots,m_u}$ and in $k$ such that $f(\vect{m},k)\neq
\gamma^kr(\vect{m},k)$ for all $r(\vect{m},k)\in\KK(\vect{m},k)$ and all roots of unity $\gamma$. Let
$S=\{s_1,\dots,s_d\}\subseteq\ZZ^d$. Consider the \pisiSE-field
$\dfield{\KK(\vect{m})(k)(t)}{\sigma}$ over $\KK(\vect{m})$ with
$\sigma(k)=k+1$ and
$\sigma(t)=\frac{f(\vect{m},k+1)}{f(\vect{m},k)}t$, and define
$f_i:=\frac{f(\vect{m}+\vect{s_i},k)}{f(\vect{m},k)}t\in\KK(\vect{m})(k)(t)$.
If there are no $c_i\in\KK(\vect{m})$ and $w\in\KK(\vect{m})(k)$
such that~\eqref{Equ:ParaDF} for $g:=w\,t$, then the following
sequences, for $r$ big enough, are algebraically independent over\footnote{Here $\KK(\vect{m})=\KK(m_1,\dots,m_u)$ is a rational function field.}
$\KK(\vect{m})(n)$:
$$S_0(n)=f(\vect{m},n),\;
S_1(n)=\sum_{k=r}^nf(\vect{m}+\vect{s_1},k),\dots,S_d(n)=\sum_{k=r}^nf(\vect{m}+\vect{s_d},k),$$
i.e., there is no
$P(x_0,x_1,\dots,x_d)\in\KK(\vect{m})(n)[x_0,x_1,\dots,x_d]^*$
with~\eqref{Equ:HypRel}.
\end{Corollary}

Moreover, if one applies Zeilbergers's creative telescoping algorithm~\cite{Zeilberger:91}, the result can be reduced to the following

\begin{Corollary}\label{Cor:Zeilberger}
Let $f(m,k)$ be a hypergeometric term in
$m$ and in $k$ such that $f(m,k)\neq
\gamma^kr(m,k)$ for all $r(m,k)\in\KK(m,k)$ and all roots of unity $\gamma$. If Zeilberger's algorithm fails to find  $c_i(m)\in\KK(m)$ and $g(m,k)$
such that
$$g(m,k+1)-g(m,k)=c_0(m)f(m,k)+\dots+c_d(m)f(m+d,k),$$
then the sequence $S_0(n)=f(m,n)$ in $n$ and the sums (for some $r$ big enough)
\begin{equation}\label{Equ:ZeilSum}
S_1(n)=\sum_{k=r}^nf(m,k),\dots,S_d(n)=\sum_{k=r}^nf(m+d,k)
\end{equation}
are algebraically independent over
$\KK(m)(n)$, i.e., there is no polynomial
$P(x_0,x_1,\dots,x_d)\in\KK(m)(n)[x_0,x_1,\dots,x_d]^*$
with~\eqref{Equ:HypRel}.
\end{Corollary}

\textsf{As a consequence, Zeilberger's algorithm finds a recurrence with minimal order for sums of the type~\eqref{Equ:ZeilSum}. Even more, it shows algebraic independence of the sums!}

\begin{Example}\label{Eample:CreaHyp}
For the Ap{\'e}ry-sum
$S(m)=\sum_{k=0}^m\binom{m}{k}^2\binom{m+k}{k},$
see~\cite{Poorten:79}, Zeilberger's algorithm finds a recurrence of order
$2$, but not smaller ones. Hence, the following sequences in $n$ are
algebraically independent over $\KK(m)(n)$:
$$\binom{m}{n}^2\binom{m+n}{n},\;\sum_{k=0}^n\binom{m}{k}^2\binom{m+k}{k},\;
\sum_{k=0}^n\binom{m+1}{k}^2\binom{m+k+1}{k}.$$
\end{Example}

The following remarks are in place.
\begin{enumerate}
\item We consider $m$ as an indeterminate; if $m$ is replaced by specific integers, the sums in~\eqref{Equ:ZeilSum} might be not well defined because of poles. In particular, $r$ might be chosen too small, or $n$ cannot be arbitrarily large.
\item Moreover, the situation might drastically change, if we consider, e.g., sums of the type
\begin{equation}\label{Equ:DefiniteHyp}
S_1(m)=\sum_{k=r}^{am+b}f(m,k),\dots,S_d(m)=\sum_{k=r}^{a(m+d)+b}f(m+d,k),
\end{equation}
for integers $a,b$. In this case, the minimal order of the corresponding recurrence might be lower.
\end{enumerate}

\begin{Example}
Consider the sum
$$S_d(m,n)=\sum_{k=0}^n(-1)^k\binom{m}{k}\binom{d\,k}{m}$$
for integers $d\geq1$. As worked out in~\cite[Sec.~4.3]{PauleSchorn:95}, Zeilberger's algorithm finds only a recurrence of order $o_d=\max(d-1,1)$. Hence the sequence $f(n)=(-1)^n\binom{m}{n}\binom{d\,n}{m}$ and the sums
$$S_d(m,n),\dots,S_d(m+o_d-1,n)$$
in $n$ are algebraically independent over $\QQ(m)(n)$. But, if we set $n=m$, the situation changes drastically. In this particular case,
$$S_d(n,n)=(-d)^n,$$
in other words, only the sequences $f(n)$ and $S_d(n,n)$, $d>1$, are algebraically independent over $\QQ$.
\end{Example}

To sum up, Zeilberger's algorithm finds, in case of existence, a recurrence with minimal order for sums of the type~\eqref{Equ:ZeilSum}. And it does not succeed in this task, if the specialization to~\eqref{Equ:DefiniteHyp} introduces additional linear recurrence relations with smaller order.

\medskip

In~\cite{Abramov:03b} a criterion is given when Zeilberger's
algorithm fails to find a creative telescoping solution for a
hypergeometric input summand $f(m,k)$. If $f(m,k)$ satisfies this
criterion, then all the sequences $h(m,n), S(m,n), S(m+1,n),\dots$
are algebraically independent over $\KK(m)$.

\begin{Example}
Since
$$f(m,k)=\frac{1}{m k+1}(-1)^k\binom{m+1}{k}\binom{2m-2k-1}{m-1}$$
satisfies Abramov's criterion, see~\cite[Exp.~2]{Abramov:03b}, it follows that the sequences $f(m,n)$ and $\{S(m+i,n)|i\geq0\}$ in $n$ with
$S(m,n)=\sum_{k=0}^nf(m,k)$ are algebraically independent over $\QQ(m)(n)$.
\end{Example}

\noindent Another criterion for transcendence is the following result inspired by \cite[Sec.~5.6]{AequalB}.

\begin{Theorem}\label{Thm:DiffHyp}
Let $f_1(k),\dots,f_d(k)$ be hypergeometric terms with the following
properties:
\begin{enumerate}
\item There is a \pisiSE-field
$\dfield{\KK(k)(t_1)\dots(t_d)}{\sigma}$ over $\KK$ with
$\sigma(k)=k+1$ and with $\sigma(t_i)=\alpha_it_i$ where
$\alpha_i:=\frac{f_i(k+1)}{f_i(k)}\in\KK(k)$ for all $1\leq i\leq d$.

\item For all $1\leq i\leq d$, $f_i(k)$ is not Gosper-summable, i.e.,
$\nexists g(k)\in\KK(k)$ with $\alpha_i\,g(k+1)-g(k)=1$.
\end{enumerate}
Then the sequences $f_1(n),\dots,f_d(n)$ together with
$S_1(n),\dots, S_d(n)$ from~\eqref{Equ:SumSet}, $r$ big enough, are
algebraically independent over $\KK(n)$, i.e., there is no
$P(x_1,\dots,x_{2d})\in\KK(n)[x_1,\dots,x_{2d}]^*$ with
\begin{equation*}
P(f_1(n),\dots,f_d(n),S_1(n),\dots,S_d(n))=0\;\;\;\forall n\geq0.
\end{equation*}
\end{Theorem}

\begin{proof}
Denote $\FF:=\KK(k)(t_1)\dots(t_d)$ and suppose that there are
$\vect{0}\neq\vecT{c_1,\dots,c_d}\in\KK^d$ and
$g\in\KK(k)[t_1,\dots,t_e]$ with~\eqref{Equ:ParaDF} where
$f_i:=t_i$. By~\cite[Cor.~2]{Karr:81}
or~\cite[Cor.~3]{Schneider:05b}, $g=\sum_{i=1}^dw_i t_i+u$ with
$w_i,u\in\KK(k)$. Plugging $g$ into~\eqref{Equ:ParaDF} and doing
coefficient comparison (the $t_i$ are transcendental!) shows that
$\sigma(w_i t_i)-w_i t_i=c_i t_i$ for all $1\leq i\leq d$. By
property~(2), $c_i=0$ for all $i$; a contradiction to the
assumption. Applying Theorem~\ref{Thm:MainSumTheorem} and choosing
an appropriate $\KK$-monomorphism proves the theorem.
\end{proof}

\begin{Example}
The sequences $\{n!,\binom{m}{n},(n+m)!, \sum_{k=1}^nk!,
\sum_{k=1}^n\binom{m}{k},\sum_{k=1}^n(k+m)!\}$ in $n$ are
algebraically independent over $\KK(m)(n)$.
\end{Example}

\noindent We note that the $q$-hypergeometric case can be handled completely analogously with our machinery.

\section{Nested sums}\label{Sec:Nested}
Most of the ideas of Section~\ref{Sec:Hyp} can be
carried over to sequences in terms of generalized d'Alembertian
extensions. E.g., in~\cite{Schneider:03} we derived for the sum
$$S(m):=\sum_{k=0}^m (1+5\,\HNumber{k}(m-2k))\binom{m}{k}^5$$
a recurrence of order $4$ with creative telescoping, but failed to
find a recurrence of smaller order. Hence
Theorem~\ref{Thm:MainSumTheorem} tells us that the sequences
\begin{equation}\label{Equ:AlgrenTrans}
\big(\binom{m}{n}\big)_{n\geq0},\;\big(H_{n}\big)_{n\geq0},\;\big(S(m,n)\big)_{n\geq0},
\dots,\;\big(S(m+3,n))_{n\geq0}
\end{equation}
in $n$ with
$$S(m,n):=\sum_{k=0}^nf(m,k)=\sum_{k=0}^n
(1+5\,\HNumber{k}(m-2k))\binom{m}{k}^5$$ are algebraically
independent over $\KK(m)(n)$. Internally, \SigmaP\ works as follows:
It constructs the \pisiSE-field $\dfield{\FF}{\sigma}$ with
$\FF:=\QQ(m)(k)(b)(h)$ from Example~\ref{Exp:DField} and designs the
$\QQ(m)$-monomorphism with

\vspace*{-0.5cm}

\begin{align*}
\ev(k,j)=j,&&
\ev(b,j)=\displaystyle\prod_{i=1}^j\frac{m+1-i}{i}=\binom{m}{j},&&
\ev(h,j)=\displaystyle\sum_{i=1}^j\frac{1}{k}=H_j.
\end{align*}

\vspace*{-0.2cm}

\noindent Then it takes
\begin{align*}
f_1&=b^5 (1+5 h (m-2 k)),&
f_2&=\tfrac{b^5
   (m+1)^5 (5 h (-2
   k+m+1)+1)}{(-k+m+1)^5},\\
f_3&=\tfrac{b^5
   (m+1)^5 (m+2)^5 (5 h (-2
   k+m+2)+1)}{(-k+m+1)^5
   (-k+m+2)^5},&
f_4&=\tfrac{b^5 (m+1)^5 (m+2)^5
   (m+3)^5 (5 h (-2 k+m+3)+1)}{(-k+m+1)^5
   (-k+m+2)^5 (-k+m+3)^5};
\end{align*}
this is motivated by the fact
$\binom{m+1}{k}=\frac{m+1}{m+1-k}\binom{m}{k}$ which shows that
$\ev(f_i,k)=f(m+i-1,k)$. Finally, \SigmaP\ proves algorithmically
that there are no $g\in\FF$ and $c_i\in\QQ(m)$
with~\eqref{Equ:ParaDF}. Hence the transcendence
of~\eqref{Equ:AlgrenTrans} follows by Theorem~\ref{Thm:MainSumTheorem}.\\
We note that the sum $S(n)=S(n,n)$ has completely different
properties: it satisfies a recurrence of order two. More precisely,
as shown in~\cite{Schneider:03} we get
$$\sum_{k=0}^n
(1+5\HNumber{k}(n-2k))\binom{n}{k}^5 =
(-1)^n\,\sum_{k=0}^n\binom{n}{k}^2\,\binom{n+k}{j}.$$

\section{A transcendence criterion for products}\label{Sec:ProdTrans}

The product version of
Theorem~\ref{Thm:ConnectionDefSumAndProperSum} is
Theorem~\ref{Thm:ConnectionProd}.

\begin{Theorem}\label{Thm:ConnectionProd} Let
$\dfield{\FF}{\sigma}$ be a difference field  with constant field
$\KK$ and $\vecT{f_1,\dots,f_d}\in(\FF^*)^d$. The following
statements are equivalent.

\begin{enumerate}
\item There do not exist $\vect{0}\neq\vecT{c_1,\dots,c_d}\in\ZZ^d$ and
$g\in\FF^*$ with
\begin{equation}\label{Equ:ParaProdDF}
\frac{\sigma(g)}{g}=f_1^{c_1}\dots f_d^{c_d}.
\end{equation}

\item There is a \piE-extension
$\dfield{\FF(t_1,\dots,t_d)}{\sigma}$ of $\dfield{\FF}{\sigma}$ with
$\sigma(t_i)=f_it_i$ for $1\leq i\leq d$.

\end{enumerate}
\end{Theorem}

\begin{proof}
Suppose that $\dfield{\FF(t_1,\dots,t_d)}{\sigma}$ is a
\piE-extension of $\dfield{\FF}{\sigma}$. Moreover, assume that
there is a $g\in\FF^*$ and
$\vect{0}\neq\vecT{c_1,\dots,c_d}\in\ZZ^d$
with~\eqref{Equ:ParaProdDF}. Let $j$ be maximal with $c_j\neq0$.
Then with $w=g t_1^{-c_1}\dots
t_{j-1}^{-c_{j-1}}\in\FF(t_1,\dots,t_{j-1})^*$ we get
$\sigma(w)=f_j^{c_j} w$, a contradiction to
Theorem~\ref{Thm:DecidePiSigmaExt}.2. Conversely, let $j$ be maximal
such that $\dfield{\FF(t_1,\dots,t_j)}{\sigma}$ is a \piE-extension
of $\dfield{\FF}{\sigma}$; suppose that $j<d$. By
Theorem~\ref{Thm:DecidePiSigmaExt}.2 there is a
$g\in\FF(t_1,\dots,t_{j-1})^*$ and $c\in\ZZ$ with $\sigma(g)=f_j^c
g$. By Theorem~\ref{Thm:FirstOrderSumExt}.2 it follows that
$g=u\,t_1^{c_1}\dots t_{j-1}^{c_{j-1}}$ with $c_i\in\ZZ$ and
$u\in\FF$; clearly $u\neq0$. Thus,
$\frac{\sigma(u)}{u}=f_1^{-c_1}\dots f_{j-1}^{-c_{j-1}}f_j^c$ which
proves the theorem.
\end{proof}

\noindent Note that the existence of a solution
of~\eqref{Equ:ParaProdDF} can be checked by Karr's
algorithm~\cite{Karr:81} if $\dfield{\FF}{\sigma}$ is a
\pisiSE-field over $\KK$. The following theorem is immediate.

\begin{Theorem}\label{Thm:ProdSeq}
Let $\dfield{\FF}{\sigma}$ be a difference field with constant field
$\KK$, let $\fct{\tau}{\FF}{\seqR}$ be a $\KK$-monomor\-phism together with an o-function, and
let $\vecT{f_1,\dots,f_d}\in(\FF^*)^d$. Then the following
statements are equivalent:
\begin{enumerate}
\item There are no $g\in\FF^*$ and
$\vect{0}\neq\vecT{c_1,\dots,c_d}\in\ZZ^d$
with~\eqref{Equ:ParaProdDF}.

\item The sequences $(S_1(n))_{n\geq0},\dots,(S_d(n))_{n\geq0}$ given by
$$S_1(n):=\sprod{k=r}n\ev(f_1,k),\dots,S_d(n):=\sprod{k=r}n\ev(f_d,k),$$
for some $r$ big enough, are algebraically independent over
$\tau(\FF)$.
\end{enumerate}
\end{Theorem}

\begin{Example}
The sequences $2^n,3^n,5^n,7^n,\dots$ over the prime numbers are
algebraically independent over $\KK(n)$;
compare~\cite[Exp.~7]{Karr:81}.
\end{Example}

\noindent The following lemma is a direct consequence
of~\cite[Thm.~4.14]{Schneider:05c}; for the rational case see
also~\cite{Abramov:02}.
\begin{Lemma}
Let $\dfield{\FF(t)}{\sigma}$ be a \pisiSE-extension of
$\dfield{\FF}{\sigma}$. Let $p_1,\dots,p_d,q_1,\dots,q_d\in\FF[t]^*$
such that $\gcd(\sigma^l(p_i),q_j)=1$ for all $l\in\ZZ$ and $1\leq
i<j\leq d$; set $f_i:=\frac{p_i}{q_i}$. Then there is no
$g\in\FF(t)^*$ and $\vecT{c_1,\dots,c_d}\in\ZZ^d$
with~\eqref{Equ:ParaProdDF}.
\end{Lemma}

\begin{Corollary}
Let $p_1(k),q_1(k)\dots,p_d(k),q_d(k)\in\KK[k]$ with
$\gcd(p_i(k+l),q_j(k))=1$ for all $i,j$ and $l\in\ZZ$. Then the
sequences
$$S_1(n):=\prod_{k=r}^n\frac{p_1(k)}{q_1(k)},\dots,S_d(n):=\prod_{k=r}^n\frac{p_d(k)}{q_d(k)},$$
for some $r$ big enough, are algebraically independent over
$\KK(n)$, i.e., there is no polynomial
$P(x_1,\dots,x_d)\in\KK(n)[x_1,\dots,x_d]^*$
with~\eqref{Equ:PolyRel}.
\end{Corollary}

\section{Conclusion}

We showed that telescoping, creative telescoping and, more generally,
parameterized telescoping can be applied to obtain a criterion to
check algebraic independence of nested sum expressions. For sums
over hypergeometric terms any implementation of Zeilberger's
algorithm can be used to check transcendence. In general, the summation package
\SigmaP\ can be applied to check algebraic independence of
indefinite nested sums and products.

Moreover, using results from summation theory one can show that
whole classes of sums are transcendental. Obviously, refinements of
summation theory should give also stronger tools to prove or
disprove transcendence of sum expressions. E.g., Peter Paule's
results~\cite{Paule:06a} enable one to predict the existence of
contiguous relations. Using these results might help to refine,
e.g., Corollary~\ref{Cor:Contiguous}.

\bigskip \noindent \textbf{Note.} A preliminary version has been presented at the 19th International Conference on Formal Power Series and Algebraic Combinatorics, Nankai University, Tianjin, China, 2007.


\providecommand{\bysame}{\leavevmode\hbox to3em{\hrulefill}\thinspace}
\providecommand{\MR}{\relax\ifhmode\unskip\space\fi MR }
\providecommand{\MRhref}[2]{%
  \href{http://www.ams.org/mathscinet-getitem?mr=#1}{#2}
}
\providecommand{\href}[2]{#2}

\end{document}